\documentclass[11pt]{IEEEtran}          
\usepackage{graphics}
\usepackage{amsmath,amssymb,amsfonts}
\usepackage{subfigure}
\usepackage{caption}

\usepackage{epstopdf}
\usepackage{amsmath,mathtools}
\begin{document}
\pagenumbering{arabic}
\title{Stochastic-Geometry Based Characterization of Aggregate Interference in  TVWS Cognitive Radio Networks}
\author{ Madhukar Deshmukh,~\IEEEmembership{Member,~IEEE}, S.M. Zafaruddin,~\IEEEmembership{Member,~IEEE}, Albena Mihovska,~\IEEEmembership{Member,~IEEE},  and Ramjee Prasad,~\IEEEmembership{Fellow,~IEEE}
		\thanks{Madhukar Deshmukh  was with Centre for Tele-Infrastrukture (CTIF), Department of Electronic Systems, Aalborg University, Denmark. Currently, 	he is  with Faculty of Engineering, Bar-Ilan University, 	Ramat Gan 52900, Israel (email: madmukar.deshmukh@biu.ac.il).}
			
			\thanks{S. M. Zafaruddin is with Faculty of Engineering, Bar-Ilan University, 	Ramat Gan 52900, Israel  (email:smzafar@biu.ac.il). }
		 	\thanks{Albena Mihovska and Ramjee Prasad are with Department of Business Development and Technology, Aarhus University, Herning, Denmark (email: amihovska@btech.au.dk, ramjee@btech.au.dk).
			 }
		 
	 }	
	
\maketitle
\begin{abstract}
In this paper,  we  characterize  the worst-case interference for a finite-area TV white space  heterogeneous network using the tools of stochastic geometry. We  derive closed-form expressions on the probability  distribution function (PDF) and an average value of the aggregate interference for various values of path loss exponent. The proposed characterization  of  the interference  is simple and can be used in improving the spectrum access techniques. Using the derived PDF, we demonstrate  the performance gain in the spectrum detection of an eigenvalue-based detector for cognitive radio networks.
\end{abstract}
\section{Introduction}
Dynamic spectrum access by  cognitive radios (CRs) is central to next-generation communication networks to bridge the scarce utilization of the spectrum by licensed services  \cite{Mitola1999}.  Efficient spectrum sensing is required to limit the interference caused by the secondary users on the primary users.  The knowledge of the interference statistics at the primary and secondary receivers can limit the interference  to the primary system by designing robust spectrum sensing mechanism. Aggregate interference in a heterogeneous secondary network is characterized by different  types of services, network topology, operational behavior, and channel fading \cite{RG}. 

 There has been  a lot of research interest in how to characterize the aggregate interference for a CR network  (see \cite{Wen2010, Babaei2008,  Ghasemi2008, Lee2012, Vijayandran2012, Kusaladharma2012, Chen2012}. In  \cite{Wen2010} and \cite{Babaei2008}, a Gaussian approximation of the aggregate interference was presented. Wen et al. \cite{Wen2010}   approximated the aggregate interference to a Gaussian random variable when the average number of nodes in the forbidden range is greater than one.  Babaei et al. \cite{Babaei2008} used the central limit theorem to show the Gaussianity of the  aggregate interference considering power control at the secondary nodes. Authors in  \cite{Ghasemi2008} \cite{Lee2012}  characterized the aggregate interference in an infinite-area network by considering a Poisson's point process for the interfering nodes. These papers presented complicated  approximations and bounds on the distribution function using  moment generating function (MGF).

Recently, aggregate interference from a finite annular region around the primary receiver has been studied using the MGF and closed form expressions were presented for different values of the path loss exponent, $\alpha$ \cite{Vijayandran2012} \cite{Kusaladharma2012}, \cite{Chen2012}. Specifically, authors in \cite{Vijayandran2012} presented the probability distribution function (PDF) of the aggregate interference in a closed-form  for  $\alpha =4$.  The tools of stochastic geometry  are becoming  useful for the analysis of wireless networks, especially the interference characterization in large wireless networks and to study the fundamental limits of networks  (see  \cite{FB}, \cite{Haenggi2009}, \cite{ElSawy2013} and the references therein).

In this paper,  we study to characterize the worst-case interference for a finite-area TV white space  heterogeneous network using the tools of stochastic geometry. We  derive  closed-form expressions on the PDF of the aggregate interference  for various values of path loss exponent, $\alpha =\{2, 3, 4, 6\}$. The derived PDF is useful  in improving the spectrum access techniques for CR networks. We also derive the expected value of the aggregate interference in terms of finite-area of the annular region. 

The paper is organized as follow: First, we present the system model in Section II. In Section III, we derive closed form expressions of the PDF and expected value of the aggregate interference. Simulation results are depicted in Section IV. Finally, Section V concludes the paper.
\section{System Model}
We consider a simplex primary system, in which the primary receivers are passive (e.g. Terrestrial TV network) and the cognitive users form a  distributed and heterogeneous ad-hoc network. The primary receivers are assumed within the coverage area described by $\mathcal A(r_{\rm dec}, r_{\rm dec})$, where $r_{\rm dec}$ is the detectability  radius of the primary receivers and  $\mathcal A$ is a general area measure on $\mathbb R^2$.     The  secondary transmitters are assumed to be distributed over the Euclidean space $\mathbb R^2$ according to the PPP $\Phi$. We define  a forbidden range from the primary network as $\mathcal A_{fi} $ =  $\mathcal A(0, r_{fi})$ for the $i$-th primary receiver.  Here, $r_{fi}$ denotes the radius of the contention region within which no secondary transmission is allowed; and we define $A_f= \bigcup_i A_{fi}$ as a protection region of the radius $r_p$ from the primary transmitter.

In this paper, we consider the worst case interference scenario for the primary as well the secondary receivers. Hence, we assume that  a victim receiver can be  located at the origin $\mathcal {O}(0,0)$ of the $\Phi$, such that the primary receiver is at a very low SINR (which can be affected severely by a nearby secondary transmitter) and the secondary receiver experience the maximum  interference from  the secondary transmitters and the primary  transmitter located at  $C_1$. Further,  we consider that there is interference from a finite-area network  of radius $r_{max}$, which contributes to the maximal aggregate interference to a victim receiver. The $r_{\rm max}$ is chosen such that the   mean interference generated by the finite secondary network at least  matches up to a factor to that from an infinite network i.e. $r\rightarrow \infty$ s.t. $1-r_{\rm max}^{2-\alpha}>1-\epsilon \Rightarrow r>\epsilon^{-1/\alpha-2}$,  where $\alpha$ is the path loss exponent.

The interference at a victim receiver located at an arbitrary location $z$ from a number of CR transmitters is:
\begin{equation}
\label{eq:interf}
I(z)=\sum_{x\in\Phi} P_{x}h_{x} l(z-x),
\end{equation}
where $P_x$ is the transmit power of the CR located at $x$, $h_x$ is the channel gain, and $l(z-x)=||z-x ||^{-\alpha}$ is the  isotropic path loss function with $\alpha$ as the path loss exponent. 

\begin{figure}[htbp]
	\begin{center}
		\includegraphics[width=\columnwidth]{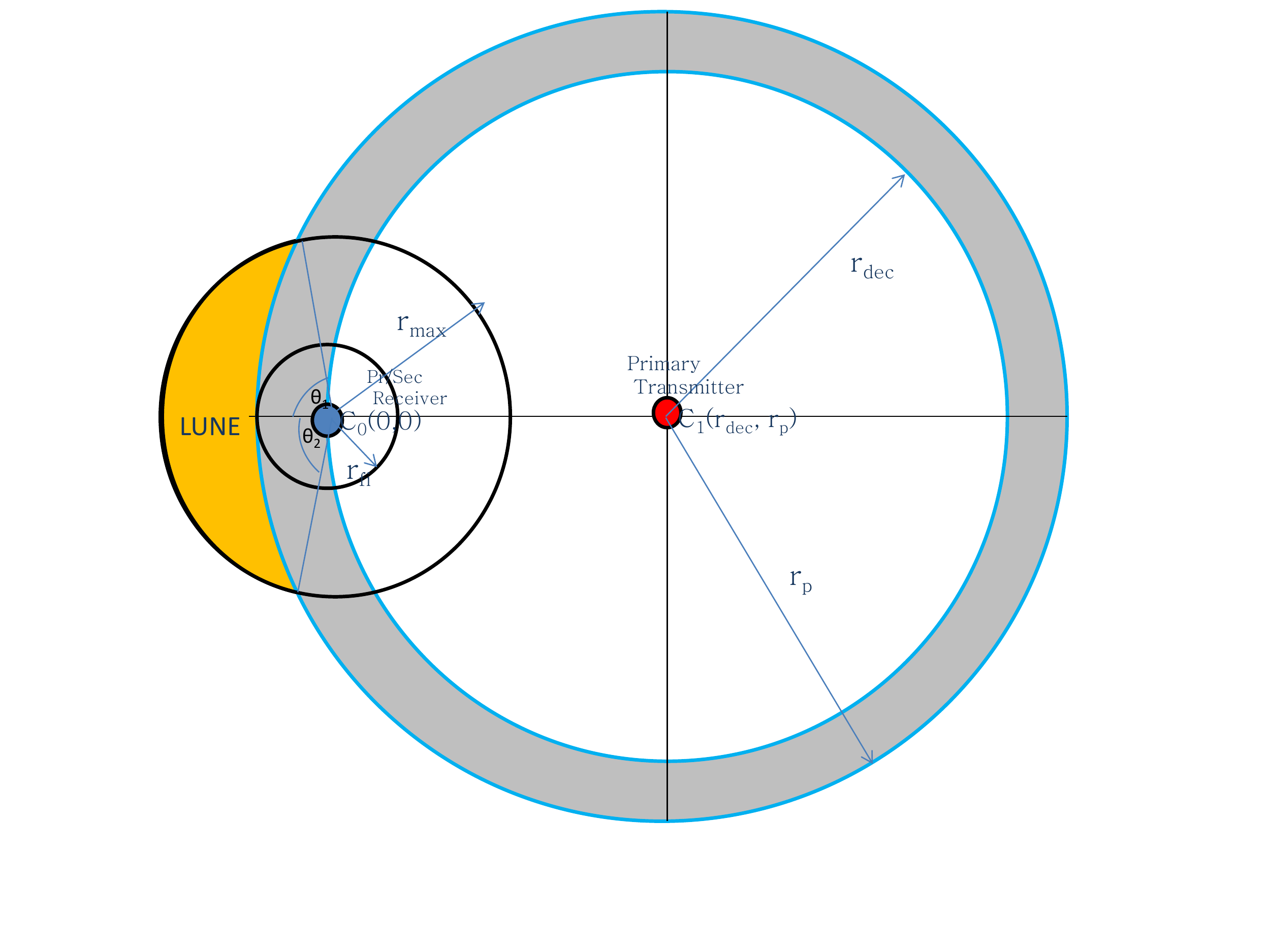}
			\caption{System model for the worst case interference.}
		\label{SysMod2}
	\end{center}
\end{figure}
\section{Statistics of Aggregate Interference}
In this section, we derive closed form expression of the PDF and expected values of aggregate interference.

\subsection{PDF of Aggregate Interference}
We use Laplace transform (LT) to characterize the interference \cite{RG} \cite{Lee2012}.  Considering the transmit power of the CRs  to unity and using $ ||x|| = ||z-x||$ , the conditional Laplace transform of the interference, $\mathcal{L}_{I(z)}^{{!o}}(s)\triangleq \mathbb{E}(e^{-sI(z)})$  in (\ref{eq:interf})  excluding the origin $\mathcal {O}(0,0)$ from the PPP, which is the location of the victim receiver:
\begin{align}
\begin{split}
\mathcal{L}^{!o}_{I(o)}(s)&=\mathbb{E}_h\big[\exp(-s\sum_{x\in\Phi\setminus \{o\}}h_{x}||x||^{-\alpha})\big]\\&=
\tilde{G}\big[\exp(-s\sum\limits_{x\epsilon\Phi\setminus \{o\}} f(x))\big]\\&=\exp\big[-\int_{\mathbb{R}^2\setminus \{o\}}(1-\exp(-sf(x))\Lambda(dx)\big]
\end{split}
\end{align}
where $\mathbb{E}_h(\cdot)$ denotes the expectation over channel realizations, $f(x)=h_{x}||x||^{-\alpha}$, $\Lambda(.)$ is the intensity function of the PPP, and $\tilde{G}(.)$ denotes the conditional probability generating functional (CPGFL) \cite{Haenggi2009}.

Since CPGFL of the PPP satisfies $\tilde{G}$(.)=$G(.)$, where $G(.)$ denotes  the CPGFL including the origin, we have
\begin{align}
\begin{split}
{\mathcal L}_{I_o}(s)&= \mathcal{L}^{!o}_{I(o)}(s)
\\&= \mathbb{E}_h~\big[\exp (-\pi\lambda{\int_{\mathbb{R}^{2}}(1-\exp(-sh_x {||x||}^{-\alpha}))dx})\big]
\end{split}
\end{align}
where ${\mathcal L}_{I_o}(s)$ is the LT of the interference including the origin, $\lambda$ denotes the intensity of the PPP,  $\mathbb{E}_h$ denotes the expectation over fading channel, and integration depicts the average over $\Phi$.

Considering the protection and forbidden region as defined in the system model, the maximal aggregate interference comes from the transmitters in the shaded region ${A_1}$, as shown in Fig. \ref{SysMod2}. First, we consider the region $A_{\rm max}$ of radius $r_{\rm max}$, and use Campbel theorem \cite{RG} to get
\begin{equation}
{\mathcal L}_{I_o}(s)= \mathbb{E}_h~\big[\exp (-\pi\lambda{\int_{0}^{r_{\max}}((1-\exp(-sh_t t^{\frac{-1}{\eta}}))dt})\big],
\label{eq:du}
\end{equation}
where $\eta=\frac{2}{\alpha}$.
Substituting $y\leftarrow t^{\frac{1}{\eta}}$ in (\ref{eq:du}), we get
\begin{equation}
{\mathcal L}_{I_o}(s)= \exp\Big(-\pi \lambda s^{\eta}\mathbb{E}(h)^{\eta}\int_{y=0}^{y=r_{\max}}y^{-\eta}e^{-y}dy\Big).
\label{INF_LT}
\end{equation}
For the infinite-area network $r_{\rm max} \to \infty$, (\ref{INF_LT}) can be calculated as \cite{RG}:
\begin{equation}
{\mathcal L}_{I_o}(s)= \exp\left(-\pi \lambda s^{\eta}\frac{\pi \eta}{sin(\pi \eta)}\right)
\end{equation}

To calculate the Laplace transform of the aggregate interference in the shaded lune  ${A_1}$, we convert the integration  in Eq. (\ref{INF_LT}) into polar form by computing the ranges of $r$ and $\theta$.  It can be easily seen that $r_p \leq r \leq r_{\rm max}$.  The angles of intersection of circle ${C_0}: u^{2}+v^{2}= r_{\rm max}^{2}$ and ${C_1}: (u-r_{dec})^{2}+v^{2}= r_{p}^{2}$ are denoted by  $\theta_1$ and $\theta_2$. Using $u=r \cos{\theta_1}$  and $v=r \sin{\theta_1}$ for $C_1$, we get $\theta_{1}$=$cos^{-1}(x/r_{\rm max})$. From Fig.~\ref{SysMod2}, it can be seen that  $\theta_{2}$=$2x-\theta_{1}$. Thus, the integral in (\ref{INF_LT}) can be represented as
\begin{equation}
T = 2\int_{\theta=0}^{\theta_1}\Big[\int_{r=0}^{r_{\rm max}}r^{-\eta}e^{-r}dr-\int_{r=0}^{r_p}r^{-\eta}e^{-r}dr\Big]d\theta
\end{equation}
Solving the above integral in terms of standard Gamma function, and substituting  in   (\ref{INF_LT}), we get
\begin{equation}
\mathcal{L}_{I_0(A_{1})} (s)=\exp(-K s^\eta),
\label{L_A1}
\end{equation}
where $K=\pi\lambda
\Gamma (1+\eta) 2\theta_1[\Gamma(1-\eta,r_{\rm max})- \Gamma(1-\eta,r_p)$.
The expression in  (\ref{L_A1}) denotes the complimentary cumulative distribution function (CCDF) of the interference which provides the worst case outage probability for both primary or secondary receivers.

Finally, we list the  PDF of the interference for various $\alpha$ by taking inverse Laplace transform  of (\ref{L_A1})
\cite{Helfand1983}:

\begin{align}
\label{eq:main}
f_I(r)=
\begin{dcases}
\delta(K), &\alpha=2 \\
\frac{2^{4/3}}{3^{3/2}\sqrt{\pi} K^{2}r^{7/3}}\exp\left(-\frac{4}{27 (K^{3} r^2)}\right)\\ \times U\left[1/6,4/3,4/27(K^{3/4} r^{-2})\right], &\alpha=3\\
\frac{1}{2 \sqrt{\pi}K r^{3/2}}\exp(-\frac{1}{(4 K^2 r)}), & \alpha=4 \\
\frac{3 }{3 K r^{4/3}}A_i(3 K r^{1/3}), & \alpha=6\\
\end{dcases}
\end{align}
where $U(.)$ is confluent hyper-geometric function and  $A_i(.)$ is the  Airy function as defined in \cite{Helfand1983}. It is important to note that  we get the same expression for PDF when $\alpha=4$ as in \cite{Vijayandran2012}.

\subsection{Expected Value  of Aggregate Interference}
The expected value of the aggregate interference $\bar{I} = \int_{0}^{r_{\rm max}} rf_I(r)dr$ over the annular region $r_{\rm max}$ for $\alpha=2$ and $\alpha=4$ is easily integrable and can be expressed as:

\begin{align}
\label{eq:main_average}
\bar{I}=
\begin{dcases}
\delta(K)\frac{r_{\max}^2}{2}, &\alpha=2 \\
{\rm erfc}(\frac{1}{2K\sqrt{r_{\rm max}}}), & \alpha=4 \\
\end{dcases}
\end{align}
where {\rm erfc} is complementary error function.

However, for $\alpha=6$, we use the asymptotic approximation of the Airy function \cite{Meissen2013}:

\begin{align}
A_i(x)\approx \frac{1}{2\sqrt{\pi}x^{1/4}} \exp(-2 x^{3/2}/3)
\label{eq:appr_airy}
\end{align}
Using the approximation on  Airy function in \eqref{eq:appr_airy}, the average of aggregate interference for $\alpha=6$ is given by

\begin{align}
\label{eq:main_average_alpha6}
\begin{split}
\bar{I}\approx \frac{1.6633r_{\rm max}^{1/9}}{K}\left( 1.3541-\Gamma(2/3, 5.1962 K^{3/2}\sqrt{r_{\rm max}}\right) \\- \frac{0.5618r_{\rm max}^{1/4}}{K^{9/4}} \Gamma(2/3, 1.5197K^{3/4}\sqrt{r_{\rm max}})
\end{split}
\end{align}

For $\alpha=3$ involving the confluent hyper-geometric function  $U(.)$, we fix the $r_{\rm max}$ to a constant value, say $3.4$ such that $K=0.5598$ and hyper-geometric function reduces to a standard gamma function. Hence, for $r_{\rm max}=3.4$, and $\alpha=3$, the average of aggregate interference is given by
\begin{align}
\label{eq:main_average_alpha3}
\bar{I}=0.56\Gamma^{-1}(1/6,0.8536r_{\rm max}^2),
\end{align}
where $\Gamma^{-1}$ denotes the inverse  Gamma function.

The expected values derived in \eqref{eq:main_average}, \eqref{eq:main_average_alpha6} and \eqref{eq:main_average_alpha3} are simple to evaluate, and can help design robust spectrum sensing techniques for CR networks. 
\begin{figure}[t]
	\begin{center}
		\includegraphics[width=\columnwidth]{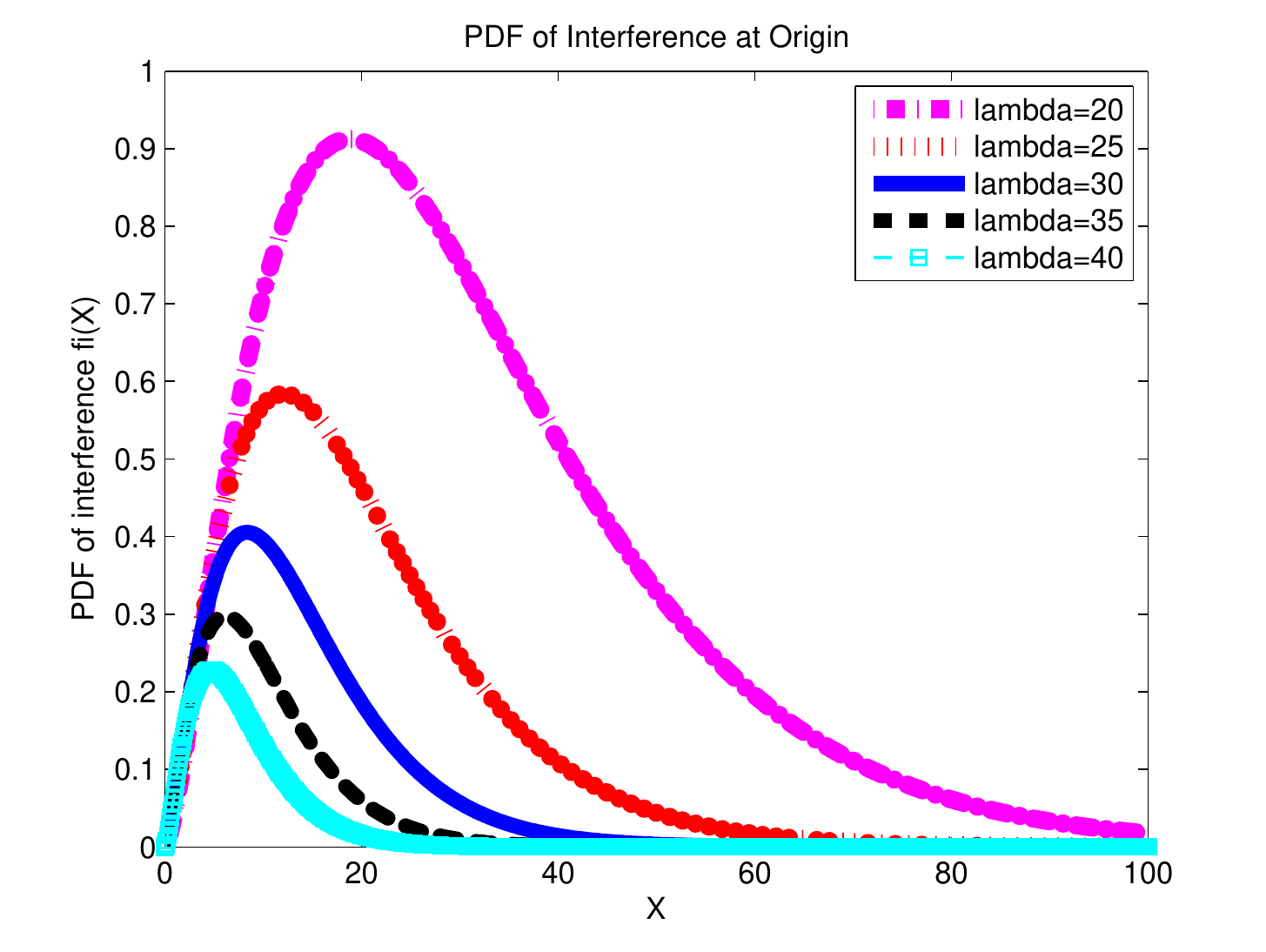}
			\caption{Probability density function of the aggregate interference for $\alpha=4$.}
		\label{pdf_inter}
	\end{center}
\end{figure}

\begin{figure}[t]
	\begin{center}
		\includegraphics[width=\columnwidth]{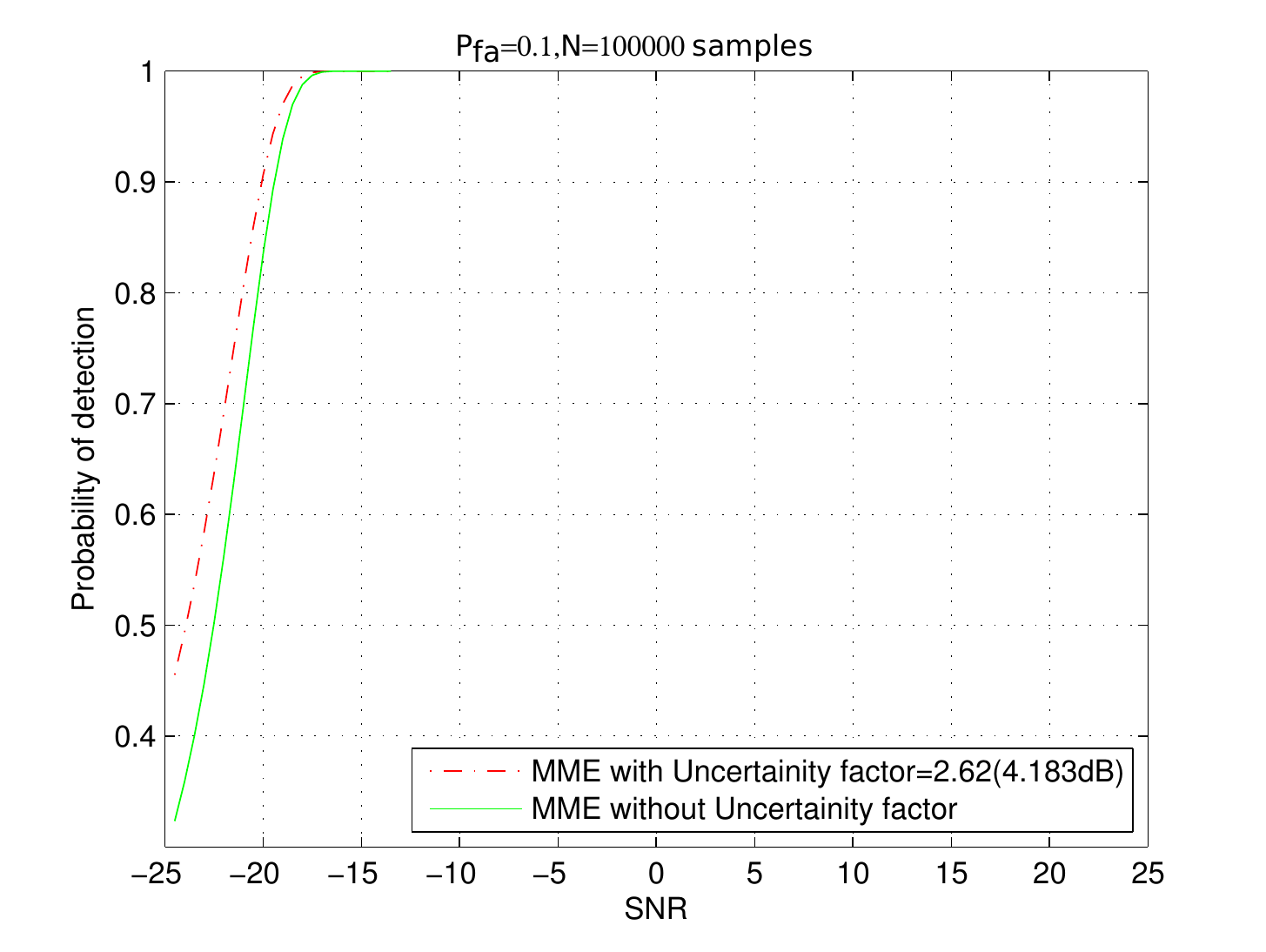}
			\caption{Impact of uncertainty on the detection probability at $\delta=2.62$, probability of false alarm $P_{fa}=0.1$, number of samples $N=100000$, and $\alpha=4$}
		\label{detect}
	\end{center}
\end{figure}

\section{Simulation Results}
Using  typical values $r_{\rm max}=3.4$,  $r_{p}=70$, $\alpha=4$ in (\ref{eq:main}), we plot the PDF of the aggregate interference in Fig.~\ref{pdf_inter}, which shows the non-Gaussian distribution of the interference.
Next, we use the derived PDF to improve the detection probability  of maximum-minimum eigenvalue (MME) \cite{Zeng2009}) spectrum detector by computing the distributional uncertainty of the aggregate interference in the form of differential entropy  \cite{Wang2008un}:
\begin{equation}
\delta=\int_{0}^{\infty} f_I(r)\ln (f_I(r))dr.
\label{uncert}
\end{equation}
Substituting  the derived PDF of \eqref{eq:main} in   (\ref{uncert}), the uncertainty in the threshold computation is improved. This improves the detection  probability of the MME spectrum detector  at low signal to noise ratio (SNR), as depicted in Fig.~\ref{detect}.


\section{Conclusion and Future Work}
We have derived closed-form expressions on  the distribution of the aggregate interference under Rayleigh fading channels for various values of path loss exponent, $\alpha =\{2, 3, 4, 6\}$.  The  derived distribution  is shown to be non-Gaussian for   the worst-case aggregate interference.  The proposed characterization  of  the interference  is simple. This can be used in improving the spectrum access techniques and  for capacity analysis in CR networks. We also derived the expected values of the aggregate interference which can helpful in designing robust spectrum sensing techniques for CR networks.

\bibliographystyle{IEEEtran}
\bibliography{15815allref}

\begin{thebibliography}{10}
\providecommand{\url}[1]{#1}
\csname url@samestyle\endcsname
\providecommand{\newblock}{\relax}
\providecommand{\bibinfo}[2]{#2}
\providecommand{\BIBentrySTDinterwordspacing}{\spaceskip=0pt\relax}
\providecommand{\BIBentryALTinterwordstretchfactor}{4}
\providecommand{\BIBentryALTinterwordspacing}{\spaceskip=\fontdimen2\font plus
\BIBentryALTinterwordstretchfactor\fontdimen3\font minus
  \fontdimen4\font\relax}
\providecommand{\BIBforeignlanguage}[2]{{%
\expandafter\ifx\csname l@#1\endcsname\relax
\typeout{** WARNING: IEEEtran.bst: No hyphenation pattern has been}%
\typeout{** loaded for the language `#1'. Using the pattern for}%
\typeout{** the default language instead.}%
\else
\language=\csname l@#1\endcsname
\fi
#2}}
\providecommand{\BIBdecl}{\relax}
\BIBdecl

\bibitem{Mitola1999}
J.~Mitola and G.~Q. Maguire, ``Cognitive radio: making software radios more
  personal,'' \emph{IEEE Personal Communications}, vol.~6, no.~4, pp. 13--18,
  Aug 1999.

\bibitem{RG}
M.~Haenggi and R.~K. Ganti, \emph{Interference in Large Wireless
  Networks}.\hskip 1em plus 0.5em minus 0.4em\relax Now Foundations and Trends,
  2009.

\bibitem{Wen2010}
Y.~Wen, S.~Loyka, and A.~Yongacoglu, ``On distribution of aggregate
  interference in cognitive radio networks,'' in \emph{2010 25th Biennial
  Symposium on Communications}, May 2010, pp. 265--268.

\bibitem{Babaei2008}
A.~Babaei and B.~Jabbari, ``Internodal distance distribution and power control
  for coexisting radio networks,'' in \emph{IEEE GLOBECOM 2008 - 2008 IEEE
  Global Telecommunications Conference}, Nov 2008, pp. 1--5.

\bibitem{Ghasemi2008}
A.~Ghasemi and E.~S. Souza, ``Interference aggregation in spectrum-sensing
  cognitive wireless networks,'' \emph{IEEE Journal of Selected Topics in
  Signal Processing}, vol.~2, no.~1, pp. 41--56, Feb 2008.

\bibitem{Lee2012}
C.~han Lee and M.~Haenggi, ``Interference and outage in poisson cognitive
  networks,'' \emph{IEEE Transactions on Wireless Communications}, vol.~11,
  no.~4, pp. 1392--1401, April 2012.

\bibitem{Vijayandran2012}
L.~Vijayandran, P.~Dharmawansa, T.~Ekman, and C.~Tellambura, ``Analysis of
  aggregate interference and primary system performance in finite area
  cognitive radio networks,'' \emph{IEEE Transactions on Communications},
  vol.~60, no.~7, pp. 1811--1822, July 2012.

\bibitem{Kusaladharma2012}
S.~Kusaladharma and C.~Tellambura, ``Aggregate interference analysis for
  underlay cognitive radio networks,'' \emph{IEEE Wireless Communications
  Letters}, vol.~1, no.~6, pp. 641--644, December 2012.

\bibitem{Chen2012}
Z.~Chen, C.-X. Wang, X.~Hong, J.~Thompson, V.~S.A., X.~Ge, X.~Hailin, and
  F.~Zhao, ``Aggregate interference modeling in cognitive radio networks with
  power and contention control,'' \emph{IEEE Transactions on Communications},
  vol.~60, no.~2, pp. 456--468, February 2012.

\bibitem{FB}
F.~Baccelli and B.~Blaszczyszyn, ``Stochastic geometry and wireless networks,''
  \emph{Applications, INRIA and Ecole normale superieure}, vol.~2, no.~45, Dec
  2009.

\bibitem{Haenggi2009}
M.~Haenggi, J.~G. Andrews, F.~Baccelli, O.~Dousse, and M.~Franceschetti,
  ``Stochastic geometry and random graphs for the analysis and design of
  wireless networks,'' \emph{IEEE Journal on Selected Areas in Communications},
  vol.~27, no.~7, pp. 1029--1046, September 2009.

\bibitem{ElSawy2013}
H.~ElSawy, E.~Hossain, and M.~Haenggi, ``Stochastic geometry for modeling,
  analysis, and design of multi-tier and cognitive cellular wireless networks:
  A survey,'' \emph{IEEE Communications Surveys Tutorials}, vol.~15, no.~3, pp.
  996--1019, Third 2013.

\bibitem{Helfand1983}
E.~Helfand, ``On inversion of the {Williams-Watts} function for large
  relaxation times,'' \emph{The Journal of Chemical Physics}, vol.~78, no.~4,
  pp. 1931--1934, Third 1983.

\bibitem{Meissen2013}
E.~Meissen, ``Probability distribution and entropy as a measure of
  uncertainty,'' \emph{Available at:math.arizona.edu/
  meissen/docs/asymptotics.pdf [online][Accessed 23 June 2018]}.

\bibitem{Zeng2009}
Y.~Zeng and Y.~C. Liang, ``Eigenvalue-based spectrum sensing algorithms for
  cognitive radio,'' \emph{IEEE Transactions on Communications}, vol.~57,
  no.~6, pp. 1784--1793, June 2009.

\bibitem{Wang2008un}
N.~Horiya and A.~Sahai, ``Probability distribution and entropy as a measure of
  uncertainty,'' \emph{Journal of Physics A: Mathematical and Theoretical},
  vol.~41, pp. 13--16, 2008.

\end{thebibliography}

\end{document}